\documentclass[fleqn]{article}
\usepackage[utf8]{inputenc}

\usepackage{amsmath}
\usepackage{mathrsfs}
\usepackage{rotating}
\usepackage{graphicx}
\usepackage{hyperref}

\makeatletter 
\renewcommand\@biblabel[1]{#1} 
\makeatother %

\title{Stochastic mechanism for improving selectivity of olfactory projection neurons}
\author{A.K.Vidybida}

\begin{document}

\begin{abstract}
\noindent
A mechanism is proposed for increasing selectivity of olfactory bulb proje\-c\-tion neurons
as compared to the olfactory receptor neurons, which could operate under low odor concentration,
when the lateral inhibition mechanism becomes inefficient. 
The mechanism proposed is based on the threshold-type reaction to stimuli a
projection neuron receives from the receptor neurons, the stochastic nature of those stimuli
and electrical leakage in the projection neurons.
The mechanism operates at the level of individual projection neuron and 
does not require involvement of other bulbar neurons.
\\
{\small\bf Keywords: } \small olfactory receptor neuron; projection neuron;
selectivity; stochastic process; theory
\end{abstract}

\maketitle

\section{Introduction}

Primary reception of odors happens in the olfactory receptor neurons (ORN).
The ORNs synapse onto mitral and tufted cells of olfactory bulb.
These cells, known as bulbar projection neurons (PN), convey odor signals to olfactory cortex.

Communication between ORNs and PNs is of convergent nature:
many ORNs synapse onto a single PN. The convergence degree depends on species and can be 
fairly large, \cite{Ressler1994}. This can ensure high sensitivity to odors,
\cite{Drongelen1978,Drongelen1978a,Duchamp-Viret1989}.

It is known that discriminating ability in PN is better than that in ORN, \cite{Duchamp1982,Kikuta2013}.
An established point of view is that better selectivity in PN is due to mechanism of lateral
inhibition, \cite{Duchamp1982,Davison2003,Linster2005}, 
which is well studied for visual system, where it increases contrast between 
domains of visual field \cite{Granit1952,Barlow1953,Hartline1956}. 
In the olfactory system, lateral inhibition is realized through granular cells, which
are stimulated by mithral cells and inhibit another PNs \cite{Yokoi1995,Urban2002}. 
As a result, the system of PNs functions in accordance to ``winner takes all'' principle,
and this can be the reason of PN having better selectivity than ORN.

In recent studies, \cite{Fantana2008}, it was realized that unlike in retina,
lateral inhibition in olfactory bulb is non-topographical.
Such a possibility was discussed before, \cite{Linster2005}.
If so, then is lateral inhibition able to ensure the same ``contrast enhancement''
in olfaction as it does in vision? This question is discussed in \cite{Valley2008}.
A final answer to this question requires additional experimental studies.

Lateral inhibition of PNs happens due to activity of inhibitory bulbar neurons. Recruitment of 
inhibitory neurons takes place for high odor concentrations and decreases with decreasing
concentration, \cite{Duchamp1982}. 
Therefore, efficacy of lateral inhibition in improving selectivity of PNs should
decrease for low concentrations. Such a decrease has been observed, 
see e.g. \cite{Duchamp-Viret1990}.

In this paper, another mechanism is proposed for selectivity gain in PNs,
which is independent of lateral inhibition and could be as well efficient
for low concentrations. This mechanism takes place for individual PN
without involvement of other bulbar cells. The prerequisites of this mechanism are as follows:
\begin{itemize}
\item[(i)] the random nature of stimuli obtained by PN from ORNs,
\item[(ii)] the threshold-type response of PN on those stimuli,
\item[(iii)] the leakage in the PN's membrane.
\end{itemize}
Similar mechanism is also possible for individual ORN, \cite{Vidybida2000,Vidybida2008b}, 
as well as in ``electronic nose'' sensors based on adsorption-desorption of odors,
 \cite{Vidybida2003a}.
 
 In this theoretical paper, as a PN model the neuronal model is used,
 which has been proposed before, \cite{Korolyuk1967}.
 Activity of single ORN is described as a Poisson process.
 Communication between a set of ORNs and corresponding PN is characterized 
 by the convergence degree, $N$, and minimal number $N_0$ of input spikes
 required for triggering that PN (firing threshold). 
 For this system, coefficient of selectivity gain, $g$,
 is defined, which shows how much a PN's selectivity is improved as compared with that of ORN.
Exact expression for $g$ as a function of system parameters is found.
 The expression's behavior is analyzed for changing parameters.
 In particular, it is observed that for physiologically relevant parameters
PN's  selectivity can be several tens times better than that of ORN.

\section{Methods}
\subsection{Neuronal model with random living time of obtained excitatory impulses}\label{Neuron}
As a model of PN the one proposed in \cite{Korolyuk1967} is used.
In this model, the membrane leakage is realized through random decay of individual input
impulses. Before the decay, each impulse is stored unchanged and disappears at
the decay moment. Thus, there is a finite set of possible values of depolarization.
The random living time of single obtained impulse has exponential distribution.
Therefore, decay of total depolarization is exponential as it should be, but
the depolarization decreases by finite jumps with height equal to the height of input impulse.
If the impulse height is small as compared with the firing threshold, then this model
satisfactorily describes membrane leakage.

Mathematically, the model can be formulated as follows.
The neuron's resting state is characterized by zero depolarization, $V=0$.
Obtaining input impulse increases depolarization by $h$, the height of input impulse.
The $h$ is analogous to EPSP amplitude.
Between the moments of obtaining two consecutive impulses, depolarization
does not change, $V(t)=const$. Therefore, at any moment of time, depolarization
takes a value from the discrete set: $V\in\{0,h,2h,\allowbreak 3h,\dots\}$.
The neuron is characterized by a firing threshold $V_0$: if depolarization is greater
than $V_0$, then the neuron generates output spike and appears in its resting state.
The triggering condition formulated in terms of $V_0$ can be reformulated in terms of the minimal number
$N_0$ of input spikes able to trigger:
$$
N_0=[V_0/h]+1,
$$
where brackets $[x]$ denote the integer part of $x$.

Until now, the model described corresponds with the model known as ``perfect integrator'',
 \cite{Abbott1999}. It is additionally expected in  \cite{Korolyuk1967} that 
 any impulse obtained by neuron has random living time.
 The living time is exponentially distributed with constant $\mu$.
 That means that any impulse may disappear during small interval $[t;t+dt[$ with probability
 $\mu\,dt$.
If, at moment $t$, the neuron keeps $k$ excitatory impulses, than depolarization is equal
to  $V(t)=kh$. Let stimulation is absent after $t$. During short interval $[t;t+dt[$,
any of $k$ impulses can decay/disappear. Expect that the impulses decay independently.
Then the probability that depolarization decreases by $h$ during $dt$ is $k\mu\,dt$.
Thus, at the end of interval  $[t;t+dt[$ one has depolarization $V(t+dt)=(k-1)h$ 
with probability $k\mu\,dt$, and $V(t+dt)=kh$ with probability $1-k\mu\,dt$.
By averaging over many realizations one has for the mean depolarization:
$$
\overline{V(t+dt)}=(k-1)hk\mu\,dt+kh(1-k\mu\,dt)=kh(1-\mu\,dt)\approx V(t)e^{-\mu\,dt}.
$$
It is clear from the latter that on the average depolarization decreases exponentially
as it should be for electrical leakage,
and constant $\mu$ has physical meaning of inverse membrane relaxation time: $\mu=1/\tau$.

\subsection{Projection neuron stimulated by many ORN}

\begin{figure}[h]
  \centerline{\includegraphics[angle=-90,width=0.4\textwidth]{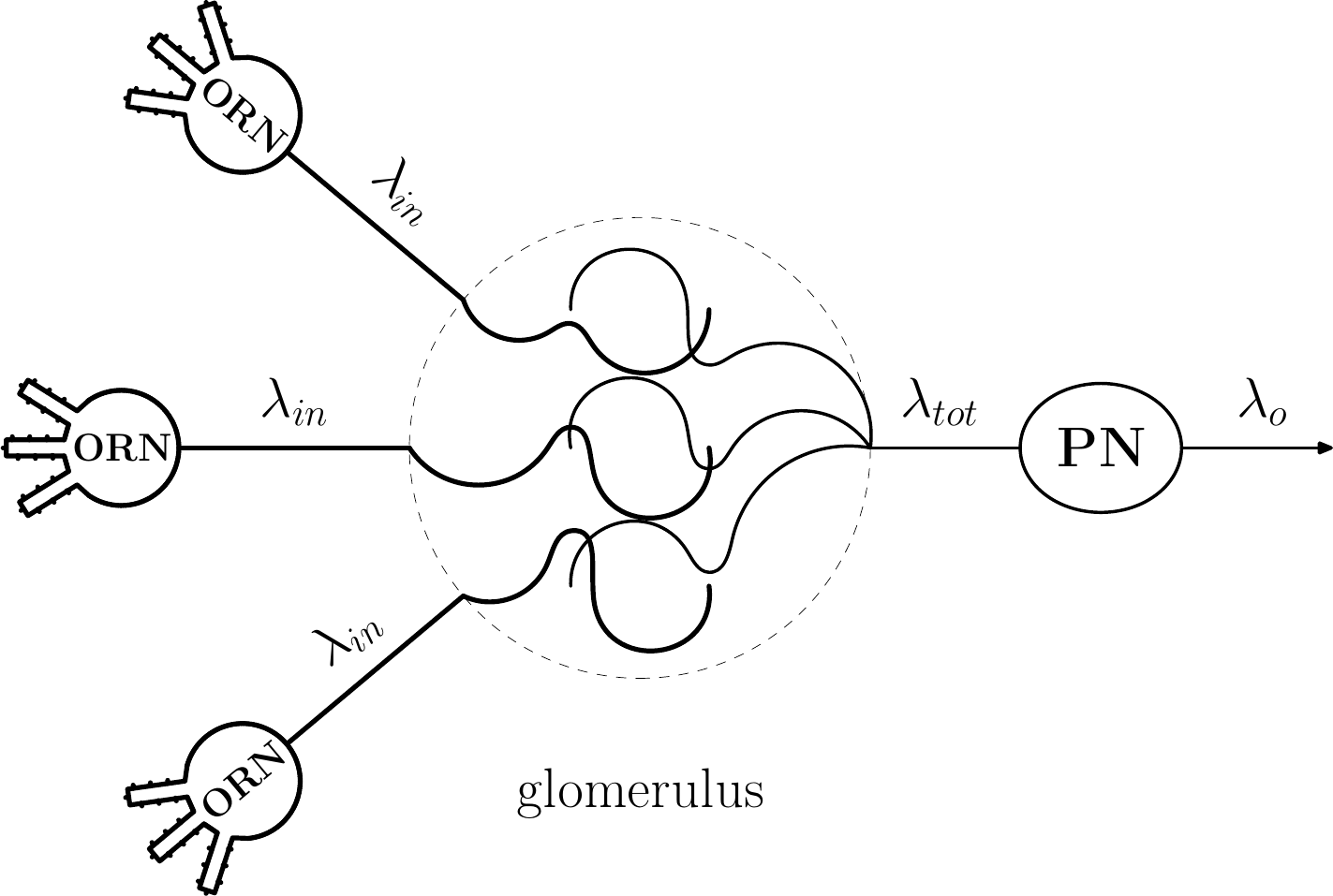}}
  \caption{\label{conver} Schematic example of communication between ORNs and PN.
  Here up to several thousands ORNs, \cite{Buck2000}, (concrete number depends on species)
  can converge through a single glomerulus onto a single PN. All those ORNs
  express the same receptor protein.}
\end{figure}

The communication scheme between ORNs and PN is shown in Fig. \ref{conver}.
It is not necessary to consider additional cells, in particular granular ones,
and additional dendrites possibly ending in other glomeruli or nearby for investigating 
how randomness, threshold and leakage influence the PN's selectivity.

Let $N$ denote the number of ORNs converging onto a single PN.
When stimulated with odor, each ORN generates a random stream of spikes
contributing to the compound stimulus applied to PN.
Taking into account specifics of primary odor reception through the receptor proteins,
it is natural to consider the ORN's output stream as a Poisson one with intensity 
$\lambda_{in}$, where subscript ``in'' indicates that the stream is considered as an
input to PN from a single ORN.
The integral input stream to the PN will be a Poisson stream with intensity
$$
\lambda_{tot}=N\lambda_{in}.
$$

\subsection{Selectivity definition}\label{sensel}

In order to compare selectivity of individual ORN with that of PN
one has to have an exact quantitative definition of those selectivities.
In order to give such a definition, consider situation when in two separate
experiments some ORN is exposed to two different odors O and O$'$ applied
at the same concentration. This will result in spiking of the ORN with
intensities $\lambda_{in}$ and $\lambda'_{in}$, correspondingly.
Assume that the odor O$'$ expresses more affinity with the ORN's receptor proteins
than does O. Then
$\lambda'_{in}>\lambda_{in}$:
$$
\lambda'_{in}=\lambda_{in}+\Delta\lambda_{in},
$$
where $\Delta\lambda_{in}>0$. 
Selectivity, or discriminating ability, $s$, between O and O$'$  of the ORN chosen 
can be defined as the following quotient:
\begin{equation}\label{s}
s=\frac{\Delta\lambda_{in}}{\lambda_{in}}.
\end{equation}
The PN this type of ORNs converge upon will as well generate more output spikes
per unite time for the odor O$'$:
$$
\lambda'_{o}=\lambda_{o}+\Delta\lambda_{o},
$$
and selectivity of the PN, $S$, between O and O$'$  can be defined
similarly:
\begin{equation}\label{S}
S=\frac{\Delta\lambda_{o}}{\lambda_{o}}.
\end{equation}
The selectivity gain, $g$, can be defined as follows:
$$
g=\frac{S}{s}.
$$
Taking into account (\ref{s}) and (\ref{S}), the latter can be represented as derivative:
\begin{equation}\label{gldef}
g=\frac{d(\log(\lambda_o))}{d(\log(\lambda_{in}))}\,.
\end{equation}
$g$ can be called coefficient of selectivity gain.
The selectivity improvement takes place if $g>1$.

\subsection{Output intensity}\label{outp}

\begin{figure}[h]
  \centerline{\includegraphics[angle=0,width=0.8\textwidth]{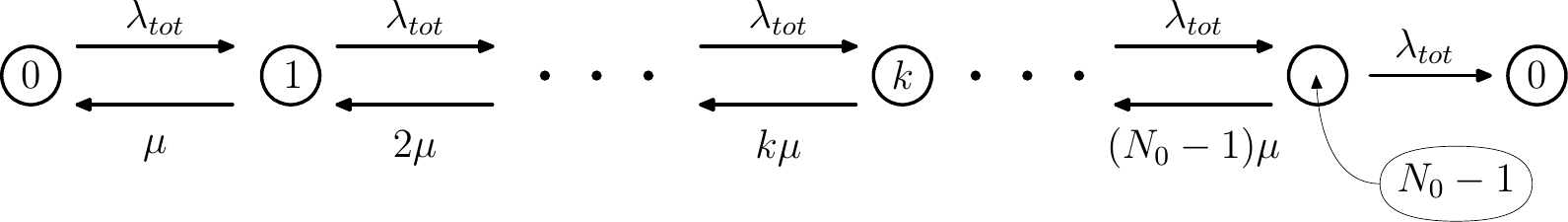}}
  \caption{\label{states}Different states of excitation of the PN. 
The inter-states transition rates are specified near the arrows. If in state number $N_0-1$
the PN obtains one more impulse, it generates a spike and appears in its resting state.}
\end{figure}

It is clear from the definition (\ref{gldef}) that in order to determine $g$ one
has to find $\lambda_o$ as a function of $\lambda_{in}$ .
Instead of output intensity $\lambda_o$ it is possible to find mean output 
inter-spike interval $T_o$. Then
\begin{equation}\label{laom0}
\lambda_o=\frac{1}{T_o}.
\end{equation}
In order to find $T_o$ consider the PN as a system with $N_0$ possible states
labeled with numbers $k=0,1,2,\dots,N_0-1$. A state with number $k$ corresponds to situation
when the PN has $k$ input impulses, see Fig. \ref{states}.
Systems of this type are known in the theory of stochastic processes as those with drain at the 
right end. A theory has been developed, which gives for a system of this kind
the mean triggering waiting time  in terms of
the transition rates and other parameters. One could use, e.g.
\cite[Eq. (1.69)]{Vidybida2006}. 
The straightforward usage of that equation with transition rates specified in the
Fig. \ref{states} results in the following expression for $T_o$:
\begin{equation}\label{m0initial}
T_o=\frac{1}{\lambda_{tot}}~\sum\limits_{0\le l\le N_0-1}~~ \sum\limits_{0\le k\le l}
\frac{l!}{k!} \left(\frac{\mu}{\lambda_{tot}}\right)^{l-k}\,.
\end{equation}

\section{Results}

\subsection{Selectivity}
After elementary transformations, instead of (\ref{m0initial}) one has the following:
\begin{equation}\label{m0next}
T_o=\frac{1}{\lambda_{tot}}~\sum\limits_{0\le j\le N_0-1}
\left(\frac{\mu}{\lambda_{tot}}\right)^{j}\frac{1}{j+1}
\frac{N_0!}{(N_0-1-j)!}.
\end{equation}

Expression for selectivity, (\ref{gldef}), with (\ref{laom0}) taken into account 
can be rewritten as follows:
$$
g=-\frac{\lambda_{in}}{T_o}\,\frac{d\,T_o}{d\,\lambda_{in}}\,.
$$
Substituting here expression for $T_o$ from (\ref{m0next}) one gets after transformations:
\begin{equation}\label{glfin}
g=
1+\frac
{\sum_{j=0}^{N_{0}-1}\frac{j}{j+1}\,\left(\frac{\mu}{N\lambda_{in}}\right)^j\,\frac{1}{(N_0-j-1)!}}
{\sum_{j=0}^{N_{0}-1}\frac{1}{j+1}\,\left(\frac{\mu}{N\lambda_{in}}\right)^j\,\frac{1}{(N_0-j-1)!}}\,.
\end{equation}
This expression is too complicated for exact analysis. In Sec. \ref{NumEx}, numerical
estimates will be given. Here, it is possible to make some limiting conclusions. 

If the PN generates an output spike in response to each input impulse from an ORN, then
 $N_0=1$. In this case, each of two sums in (\ref{glfin}) reduces to a single term with $j=0$, 
which gives no selectivity gain:
$$
N_0=1\quad\Rightarrow\quad g=1.
$$

If more then one input impulse is required for triggering, then $N_0>1$ and
(\ref{glfin}) can be presented as follows:
\begin{equation}\label{glfinfin}
g=
1+\frac
{\sum_{j=1}^{N_{0}-1}\frac{j}{j+1}\,\left(\frac{\mu}{N\lambda_{in}}\right)^j\,\frac{1}{(N_0-j-1)!}}
{\frac{1}{(N_0-1)!}+\sum_{j=1}^{N_{0}-1}\frac{1}{j+1}\,\left(\frac{\mu}{N\lambda_{in}}\right)^j\,\frac{1}{(N_0-j-1)!}}
\end{equation}

Consider the perfect integrator case, ($\mu=0$). 
In this case, there is no leakage and, as it can be seen from (\ref{glfinfin}), 
there is no selectivity gain:
$$
\mu=0\quad\Rightarrow\quad g=1.
$$

Similarly, there is no selectivity gain for very intensive stimulation:
$$
\lambda_{in}\to\infty\quad\Rightarrow\quad g\approx 1.
$$

If the stimulation is very weak ($\lambda_{in}\to 0$), the right-hand side of Eq. (\ref{glfinfin})
turns into $N_0$. Thus, the selectivity gain equals to the threshold height measured in the units
of the amplitude of input impulses:
$$
\lambda_{in}\to 0\quad\Rightarrow\quad g\approx N_0.
$$
The latter example should be taken with caution since under weak stimulation the output spikes
will happen too rare to have a physiological meaning.

It is possible to prove that the derivative of right-hand side of (\ref{glfin}) with respect to 
$\lambda_{in}$ is negative. Thus, $g$ decreases with increasing $\lambda_{in}$. 
Therefore, in all other cases, namely, with $N_0>1$, $\mu>0$, and a moderate stimulation applied,
the selectivity gain will be in the following limits: $1<g<N_0$, and a certain selectivity improvement
will be observed. Numerical examples are given in the Sec. \ref{NumEx}.

\subsection{Numerical examples}\label{NumEx}
In order to present numerical examples we have to chose values for quantities which appear
in the Eq. (\ref{glfinfin}). Experimental data for figuring all required quantities
for a single species are absent. Therefore, approximate estimates and analogy are used.
The quantities used for calculations are given in the Tab. \ref{T1}.
The EPSP amplitude, $h$  produced in the PN by a single spike from a ORN seems not reported.
This value is required to determine the PN's triggering threshold in the $h$ units:
$N_0\approx V_0/h$. Therefore, the $h$ for CA1 pyramidal neurons is used. 
Actual amplitude for the PN may be substantially lower due to mutual dendrite shunting 
through the gap junctions, \cite{Bourne2017,Kikuta2013}. 
The electrical leakage rate (the decay rate of input impulses, Sec. \ref{Neuron}, above)
is calculated as $\mu=\tau^{-1}=0.0111$ ms$^{-1}$.
The convergence degree is taken $N=5000$, \cite{Ressler1994}, for all cases. 
Resulting selectivity gain, $g$, and corresponding output rate, $\lambda_o$
are given in Tab. (\ref{T2}).

\begin{table}
\begin{tabular}{cccc}
\hline\\
threshold        & height      &ORN spikes   &PN membrane \\
depolarization, & of EPSP,     &frequency,   &relaxation time,   \\
$V_0$, mV      & $h$, $\mu$V  &$\lambda_{in}$, 1/ms &$\tau$, ms  \\
\hline
 5 - 12, \cite{Urban2002,Burton2014} & 30 - 665,   & 10$^{-3}$,  \cite{Tan2010} & 90, \cite{Mori1981}  \\
 & the mean is 131, \cite{Sayer1990}  & &    \\
\hline
\end{tabular}
\caption{\label{T1}Experimental values for parameters, sources are indicated in brackets.}
\end{table}

\begin{table}
\begin{tabular}{ccc}
\hline\\
               &output  &\\
threshold         &frequency      &\\             
$N_0$         &$\lambda_o$, 1/s  &$g$\\
\hline
300   & 10.3 & 1.78\\
400   & 5.3 & 3.15\\
500   & 0.67 &30.3 \\
\hline
\end{tabular}
\caption{\label{T2}Results of numerical calculations. $\lambda_o$, $g$ are calculated using Eqs.
(\ref{laom0}), (\ref{glfin}), respectfully.  $N_0$ is chosen in accordance with the data of
Table \ref{T1}.}.
\end{table}

\begin{figure}
  \centerline{\includegraphics[width=0.64\textwidth]{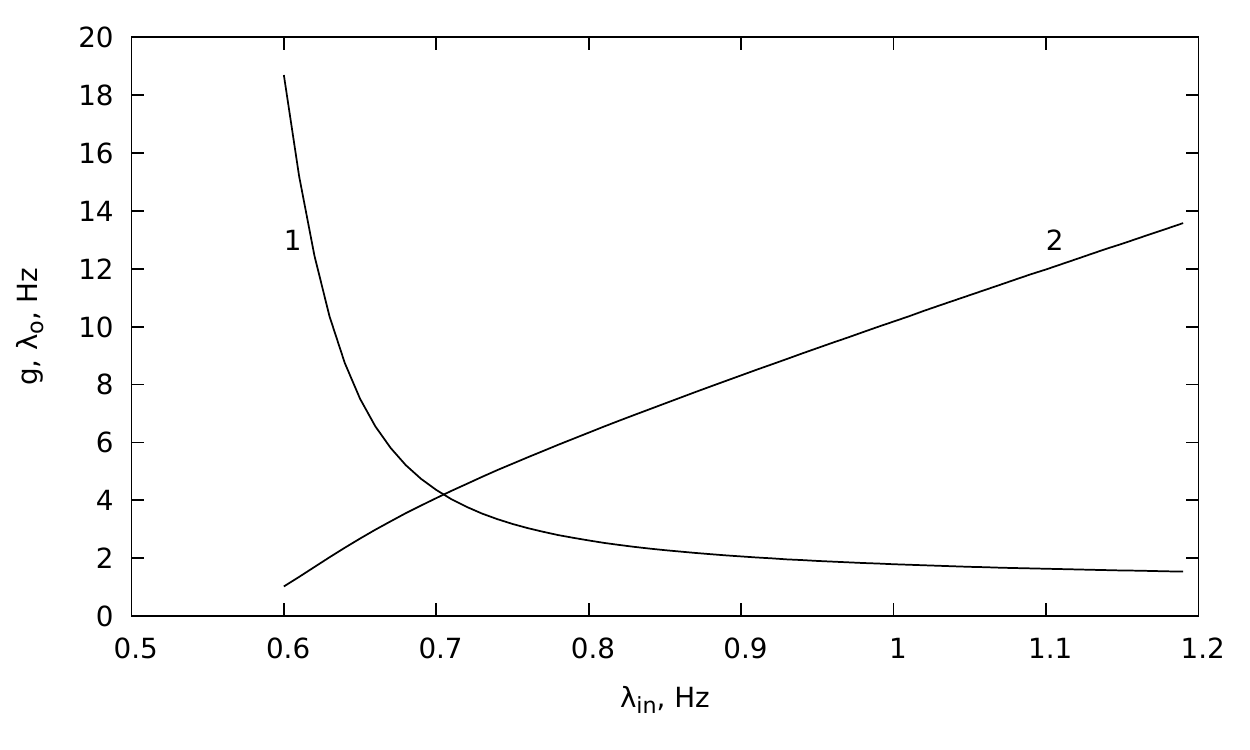}}
  \caption{\label{N_0300}Dependencies of $g$, 1 and $\lambda_o$, 2 on $\lambda_{in}$ for threshold
   $N_0=300$, $N=5000$, $\tau=90$ ms. $g$  is dimensionless.}
\end{figure}

\begin{figure}
  \centerline{\includegraphics[width=0.64\textwidth]{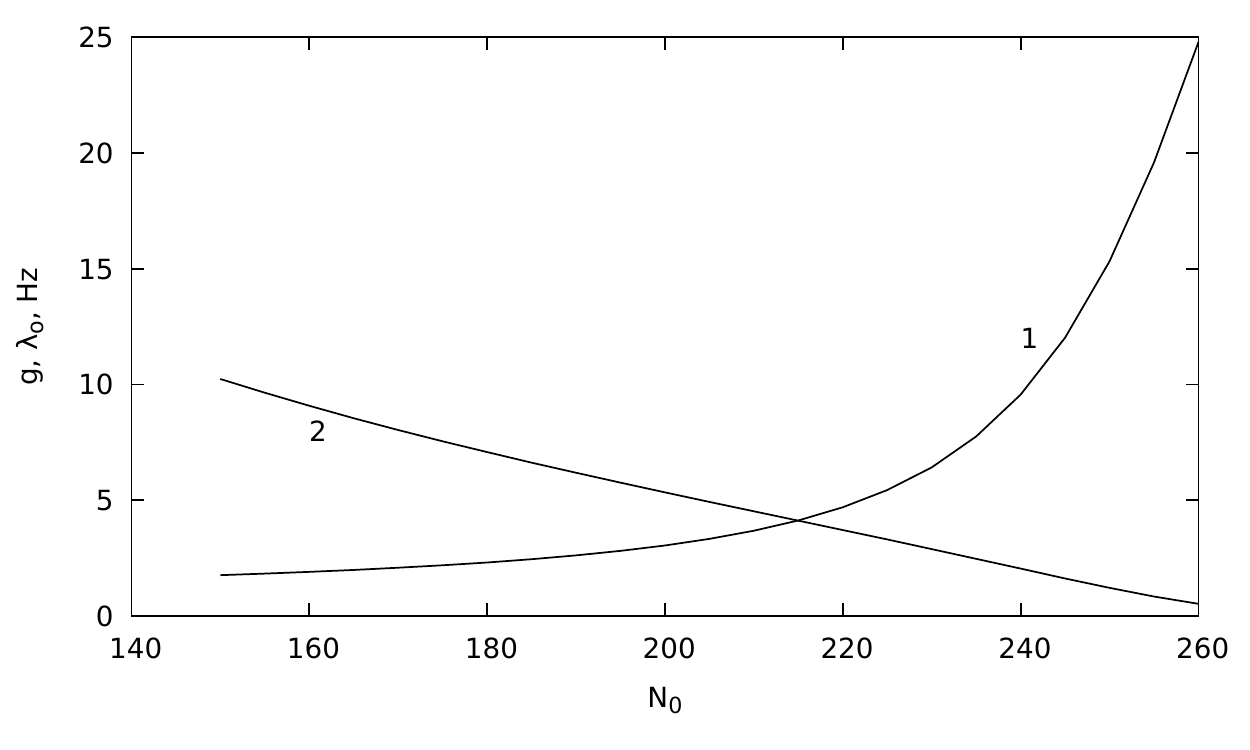}}
  \caption{\label{lai00005}Dependencies of $g$, 1 and  $\lambda_o$, 2 on threshold $N_0$ for $\lambda_{in}=0.5$ Hz.}
\end{figure}

Dependence of the desired quantities on the stimulus intensity and threshold are shown in
Figs. \ref{N_0300} and \ref{lai00005}, respectfully.

\section{Conclusions and discussion}

Higher selectivity of the secondary olfactory sensory neurons as compared to the primary
ones has been discussed regularly, 
 \cite{Duchamp1982,Duchamp1984,Duchamp-Viret1989,Duchamp-Viret1990,Duchamp-Viret1997}.
The lateral inhibition has been proposed as a sole mechanism explaining higher selectivity of PNs,
 \cite{Duchamp1982,Yokoi1995,Davison2003}.

In this paper a different mechanism is proposed, which is based exclusively on the stochastic
nature of the stimuli received by PNs, the threshold-type reaction to those stimuli
and electrical leakage in the PN's membrane.
This mechanism does not depend on lateral inhibition and is able to function at low odor concentrations. 
A coefficient of selectivity gain, $g$ is defined in order to get quantitative description.
Possible values of $g$ for physiologically real parameters are obtained.
The coefficient of selectivity gain is characterized by the following.
There is no selectivity gain if the secondary neuron is triggered by each single input impulse, $g=1$.
The selectivity gain increases with increasing triggering threshold, $N_0$.
Also, there is no gain if electrical leakage is absent. Similar situation takes place if
input stimulation is very high, when in spite of the leakage every $N_0$ input impulses trigger
the secondary neuron.
For very low intensity of input stimulation (low concentration of odor) $g$ approaches its maximal value,  $g\approx N_0$. Under moderate odor concentration $1<g<N_0$. 
For parameters taken from physiological range, the mechanism proposed can ensure several tens
times better selectivity of secondary neuron as compared with that of the primary ones.

Earlier, an idea has been proposed that the convergent nature of communication between ORNs
and PN as well might improve PN's selectivity, \cite{Drongelen1978a}. 
As to my knowledge, no physical mechanism for such an improvement has been proposed.
The mechanism proposed in this paper also does not base on conver\-gen\-ce.
Indeed, it can be seen from Eqs. (\ref{glfin}), (\ref{glfinfin}) that the degree of convergence $N$
is only used to calculate the intensity of compound stimulation from the set of ORNs, 
$\lambda_{tot}=N\lambda_{in}$.
The same value of $\lambda_{tot}$ can be ensured either by large number of low-active ORNs,
or small number of high-active ones. In both cases the same selectivity gain will be obtained
 provided other parameters are the same.
Here, it should be mentioned that spontaneous activity of ORNs has been excluded from
consideration. This activity can worsen detection of weak olfactory stimuli, \cite{Lowe1995}. 
At the same time, due to high degree of convergence, uncorrelated spontaneous noise can be
averaged out, \cite{Drongelen1978a,Linster2005}.
Therefore, convergence plays indirectly some role in the mechanism proposed.

To finish with the ORN's spontaneous activity, it should be mentioned
that that can be quite low, \cite{Rospars1994,Tan2010},
whereas the time required for odor perception can be quite short, \cite{Bhandawat2010a},
actually, much shorter than the mean interspike  interval in spontaneous activity.
Thus, it may so happen that, during odor perception time, influence of spontaneous activity
is minimal.
 
An interesting feature to discuss is the selectivity dependence on the odor concentration.
With the concentration increasing, the ORN's spiking frequency $\lambda_{in}$ increases as well.
In this case, the proposed mechanism predicts a decrease of PN's selectivity.
This is in concordance with some experimental observations,  \cite{Tan2010}.
In other experiments, it was observed either selectivity increase with
odor concentration increase, \cite{Duchamp-Viret1990}, or it was
independent of concentration, \cite{Duchamp1984}.
This contradiction could be resolved if odor concentration applied in \cite{Tan2010}
is lower than that in \cite{Duchamp-Viret1989,Duchamp-Viret1990}.
Indeed, in \cite{Duchamp-Viret1989,Duchamp-Viret1990} also a progressive recruitment 
of bulbar neurons with increasing odor concentration is observed. In this process,
the number of active inhibitory neurons grows faster than that of excitatory ones, \cite{Duchamp1982,Rospars1994}. This is a prerequisite for lateral inhibition.
The latter, at higher concentrations could be more efficient in improving PN's selectivity.
This explains selectivity increase with increasing odor concentration, observed in
\cite{Duchamp-Viret1990}. 
At the lowest concentrations, proportion of inhibitory neurons among all active in
the olfactory bulb is considerably lower than at the high ones, or inhibitory activity
is absent at all, \cite{Duchamp1982}. In this case, the lateral inhibition does not
work, whereas mechanism discussed here predicts selectivity improvement with decreasing
concentration.

It should be mentioned that the possible mechanism for selectivity gain is proposed 
here based on theoretical analyses of a considerably simplified picture.
In particular, the model used to describe a projection neuron corresponds to the
widely used leaky integrate-and-fire one only on the average. This model was proposed 
in \cite{Korolyuk1967}. It is used here because for this model it is possible to obtain
exact mathematical expressions characterizing stochastic triggering process.
This mo\-del can be a suitable approximation of the leaky integrate-and-fire model.
At the same time, processing of input spikes could happen at the level of dendritic tree, \cite{Husser2000,London2005}.
This fact can be taken into account in the current approach, but requires a sizable extension
and can be done in further publications.
Another simplification is that individual ORNs are considered as identical,
whereas they can differ from each other in sensitivity and response speed, \cite{Rospars2014}.
When estimating an ORN's activity (firing rate $\lambda_{in}$) it is not accounted that 
communication from ORN to PN is inhibited presynaptically, \cite{Mcgann2013}. 
This may result in decreasing ORN's effective activity and corresponding increase of
the firing threshold $N_0$. Also, axon from single ORN arborizes and forms several synapses 
\cite{Lledo2005a}. This could increase ORN's effective activity and cause a corresponding 
decrease of the firing threshold $N_0$.
\bigskip

{{\small\bf Acknoledgements.} \small The numerical calculations have been made by means of
free computer algebra system Maxima, \verb-http://maxima.sourceforge.net-.
This paper was supported by the 
Programs "Structure and Dynamics of Statistical and Quantum-Field Systems", 
Project PK  0117U000240 
and "Dynamics of formation of spatially non-uniform structures in many-particle systems", 
Project PK  0107U006886 of the National Academy of Science of Ukraine.

\bibliographystyle{vancouver}

\bibliography{../References}{}


\end{document}